\documentstyle[sprocl,epsfig]{article}

\def\Journal#1#2#3#4{{#1} {\bf #2}, #3 (#4)}

\def\NPA{{ Nucl. Phys. A} }
\def\PLB{{ Phys. Lett.}  B}

\def\PRL{ Phys. Rev. Lett.}

\def\PRD{{ Phys. Rev.} D}
\def\PRC{{ Phys. Rev.} C}

\def\EPJC{{Eur. Phys. J.} C}
\def\JPG{J. Phys. G}

\def\ra{\rightarrow}

\def\be{\begin{equation}}
\def\ee{\end{equation}}
\def\bea{\begin{eqnarray}}
\def\eea{\end{eqnarray}}

\def\ua{\uparrow}
\def\da{\downarrow}

\def\ubar{{\bar u}}
\def\dbar{{\bar d}}

\begin{document}

\title{FLAVOUR SYMMETRY BREAKING IN THE POLARIZED NUCLEON SEA}

\author{FU-GUANG CAO and A. I. SIGNAL}

\address{Institute of Fundamental Sciences, Massey University \\
Palmerston North, New Zealand
}

\maketitle\abstracts{
After a brief review on flavour symmetry breaking (FSB) in the unpolarized
nucleon sea,
we discuss theoretical predications for FSB in the polarized nucleon sea
from meson cloud and `Pauli blocking'.
}

\section{Flavour symmetry breaking in the unpolarized nucleon sea}
The possible breaking of parton model symmetries by the nucleon's quark 
distribution functions has been a topic of great interest since the experimental 
discoveries that the Ellis-Jaffe and Gottfried sum rules 
are violated. In particular, the flavour asymmetry in the nucleon sea 
($\dbar > \ubar$) has been confirmed by several experiments \cite{NA51+E866}.
This asymmetry can be naturally explained in the meson cloud model \cite{Thomas83},
in which the physical nucleon can be viewed as a bare nucleon
plus some meson-baryon Fock states which result from the fluctuation
$N \ra M B$,
\bea
|p\rangle_{\rm phys.}= |p\rangle_{\rm bare} + |\pi N\rangle
 + |\pi \Delta\rangle + \cdots .
\label{NMCM}
\eea
The valence anti-quark in the meson contributes (via a convolution)  
to the anti-quark distributions in the proton sea.
Since the probability of the Fock state $|n\pi^+\rangle$  is larger than that of the
$|\Delta^{++}\pi^-\rangle$ state in the proton wave function, the asymmetry 
$\dbar >\ubar$ emerges naturally in the proton sea.
Another possible source for this asymmetry is that
the bare nucleon, $|p\rangle_{\rm bare}$ may have an intrinsic
asymmetry associated with it.
According to the Pauli exclusion principle,
the $d \dbar$ is more likely to be created than the $u\ubar$ pair
since there are two valence $u$ quarks and only one
valence $d$ quark in the proton.
So there is a small excess of $d\dbar$ pairs over $u\ubar$ pairs.
This asymmetry has also been studied
in chiral quark model \cite{CQM},
the chiral quark-soliton model \cite{CQSM}, and
the instanton model \cite{Instanton}.
It was shown by Melnitchouk, Speth and Thomas
\cite{WMelnitchoukST} that
by using the meson cloud model together with the Pauli blocking,
the data for both the ratio $\dbar(x)/\ubar(x)$ and
difference $\dbar(x)-\ubar(x)$ can be described
reasonably well, while using one of these effects will not (see Fig. 1).
About half of the asymmetry can be attributed to the meson cloud
and the other half to the Pauli blocking.
We would like to point out that the data that were compared with is 
the E866 data in 1998.
Recently E866 collaboration reported its improved measurements \cite{E866_2001}
in which the statistics are improved
and the measured $x$-range is extended to lower $x$ (from $0.036$ to $0.026$).
An interesting point is that comparing with the previous results,
the ratio for $x$ being $0.315$ is pushed down from about $0.9$
to about $0.4$ and the difference $\dbar-\ubar$ becomes negative,
while the other data remain nearly unchanged.
So for the last data point in large $x$ in Fig. 1 the theoretical predications will be
well outside the error bars.
More studies are needed in theoretical calculations
and experimental measurements.
\begin{figure}		
\begin{center}
\bea
{\epsfig{figure=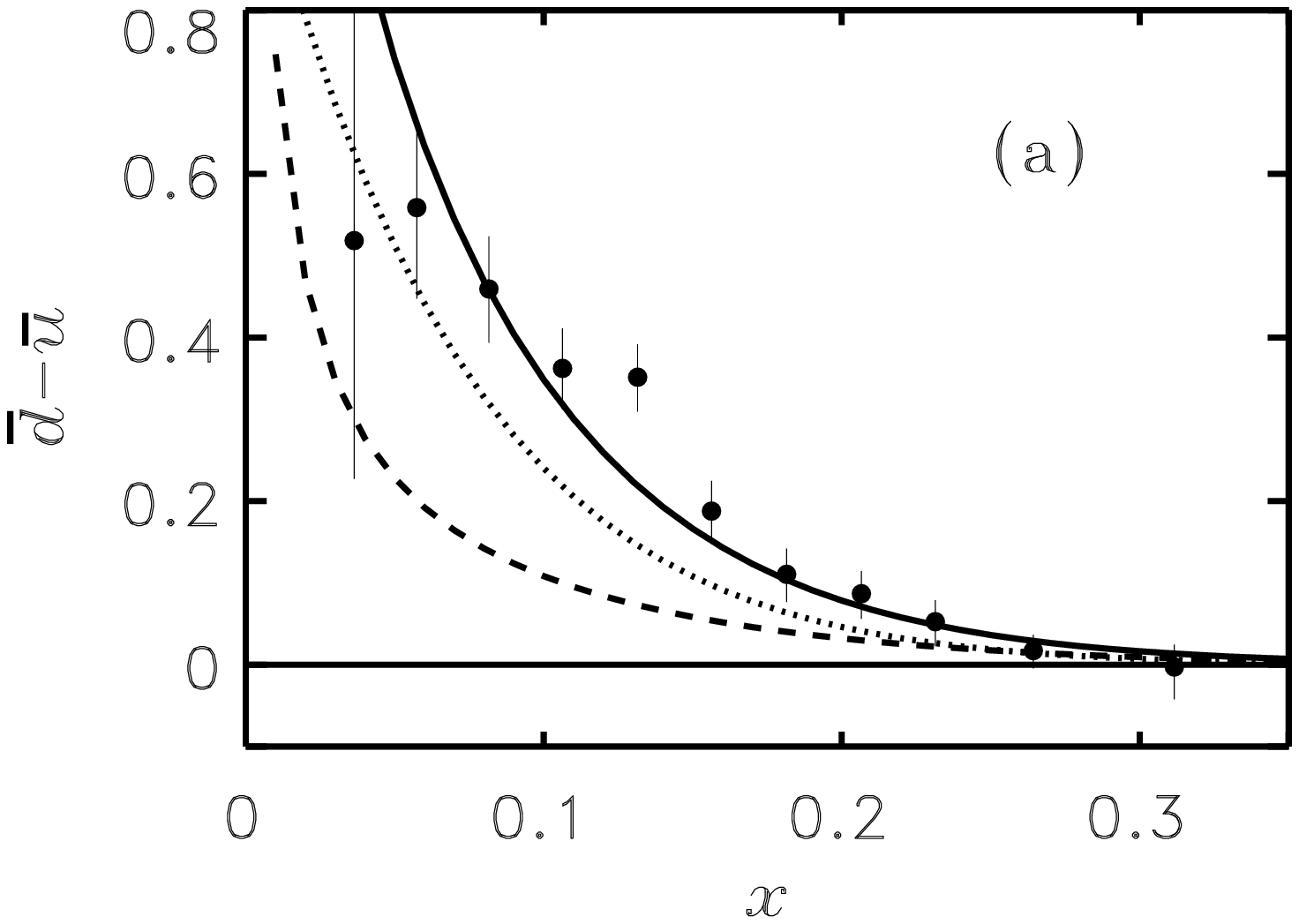,height=4cm,width=5.8cm}}
& ~~&
{\epsfig{figure=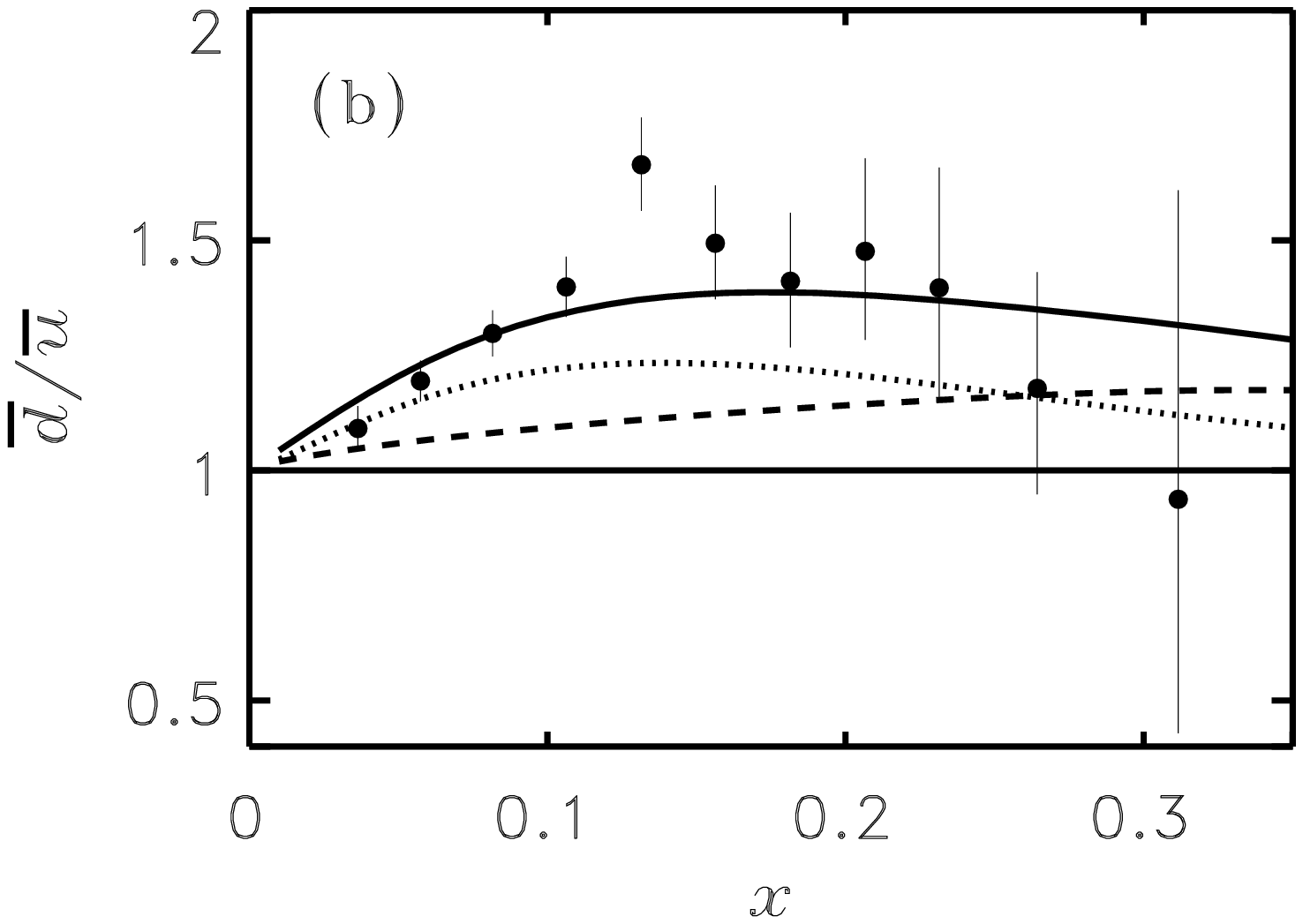,height=4cm,width=5.8cm}} \nonumber
\eea
\vspace{-0.8cm}
\caption{Contributions from pions with $\Lambda_{\pi N} = 1$~GeV
	and $\Lambda_{\pi\Delta} = 1.3$~GeV (dashed) and from
	antisymmetrization (dotted) to the
	(a) $\bar d - \bar u$ difference and
	(b) $\bar d/\bar u$ ratio,
	and the combined effect (solid). Taken from Phys. Rev. D {\bf 59}, 014003 (1998).
	 }
\end{center} 
\end{figure}

\section{Flavour symmetry breaking in the polarized nucleon sea}

Recently there has been increasing interest in the question of whether this 
asymmetry extends also to the polarized sea distributions 
{\it i.e.} $\Delta \dbar(x) \neq \Delta\ubar(x)$?
Such a polarized sea asymmetry would make a direct contribution to the Bjorken 
sum rule.
Although well established experimental evidence for a polarized sea asymmetry
is still lacking, some experimental studies have been done \cite{PAsymmetryExp}.
Moreover several parameterizations \cite{PPDFit} for the polarized parton 
distributions arising from fits of the world data from polarized experiments leave 
open the possibility of this asymmetry.
There have also been some theoretical studies on this asymmetry.
A much larger asymmetry in the polarized sea distributions than
that in the unpolarized sea distributions is predicted
in the chiral quark-soliton model
(using the large-$N_C$ limit) \cite{LargeNC}.
Such sizeable asymmetries would make an important contribution (around 20\%) to 
the Bjorken sum rule. 
This asymmetry has also been studied by considering the $\rho$ meson cloud
in the meson cloud model \cite{RFriesS}.
The prediction for  $\Delta \dbar(x) - \Delta \ubar(x)$ is more than one order of 
magnitude smaller than the result from the chiral quark-soliton model.
More theoretical calculations can be found in reference \cite{More}.
Here we report a study \cite{FGCaoS} on the flavour asymmetry of
the non-strange polarized 
anti-quarks using the meson cloud model and `Pauli blocking'.

\subsection{FSB in the meson cloud model}

It was assumed in the meson cloud model (MCM)
that the lifetime of a virtual baryon-meson Fock state is much
larger than the interaction time in the deep inelastic or Drell-Yan
process, thus the quark and anti-quark in the virtual meson-baryon Fock states
can contribute to the parton distributions of the nucleon.
For polarised parton distributions in the model it is necessary to 
include all the terms which can lead to the same final state \cite{SchT}. 
This allows the possibility of interference terms between different terms in the 
nucleon wavefunction Eq.~(\ref{NMCM}). 
For polarised anti-quark distributions the interference will be between terms 
with different mesons and the same baryon e.g. $N\pi$ and $N\rho$
(see Fig.~2).
\begin{figure}
  \begin{center}
\epsfig{file=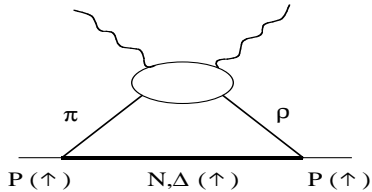,height=2.5cm,width=5cm,clip=,angle=0}
 \end{center}                                                                     
\caption{Schematic illustration of interference contributions to the polarized
anti-quark distributions.}
\end{figure}
We will consider the fluctuations $p \ra N \pi, N \rho, N \omega$ and 
$p \ra \Delta \pi, \Delta \rho$.
The fluctuation $p \ra \Delta \omega$ is neglected because of isospin conservation.
The flavour asymmetry is studied by calculating the difference
$x(\Delta\dbar -\Delta \ubar)$ which turns out to be
\bea
x(\Delta\dbar -\Delta \ubar)&=&~~[\frac{2}{3} \Delta f_{\rho N/ N}
-\frac{1}{3} \Delta f_{\rho \Delta/ N }] \otimes\Delta v_\rho \nonumber\\
&~~&+[-\Delta f_{(\rho^0\omega)p/p}+\frac{2}{3} f^0_{(\pi\rho)N/N}
-\frac{1}{3} f^0_{(\pi\rho)\Delta/N}-f^0_{(\pi^0\omega)p/p}]
\otimes\Delta v_\rho \nonumber \\
&=&\Delta f_\rho\otimes\Delta v_\rho +
	\Delta f_{int}\otimes\Delta v_\rho.
\label{xDeltadu}
\eea
where $\Delta v_\rho$ is the polarized parton distribution
of the $\rho$ meson,
$\Delta f_{(V_1 V_2)B/N} =f^1_{(V_1 V_2)B/N}-f^{-1}_{(V_1 V_2)B/N}$
is the polarized fluctuation function and
$f^\lambda_{(M_1 M_2)B/N}=\sum_{\lambda^\prime}
\int^\infty_0 d k_\perp^2
\phi^{\lambda \lambda^\prime}_{M_1 B}(y, k_\perp^2)
\phi^{*\,\lambda \lambda^\prime}_{M_2 B}(y, k_\perp^2)
$
is the helicity dependent fluctuation function.
$\phi^{\lambda \lambda^\prime}_{MB}(y,k_\perp^2)$ 
is the wave function of the Fock state containing a meson ($M$)
with longitudinal momentum fraction $y$, transverse momentum ${\bf k}_\perp$,
and helicity $\lambda$,
and a baryon ($B$) with momentum fraction $1-y$,
transverse momentum $-{\bf k}_\perp$, and helicity $\lambda^\prime$.
The first term in Eq.~(\ref{xDeltadu}) is the same as the result given in \cite{RFriesS}.
The second term  in Eq.~(\ref{xDeltadu}) is the interference contribution.
We note that there  are no contributions directly from the $\omega$ meson
due to its charge structure.

\begin{figure}
 \begin{center}
\bea
\epsfig{file=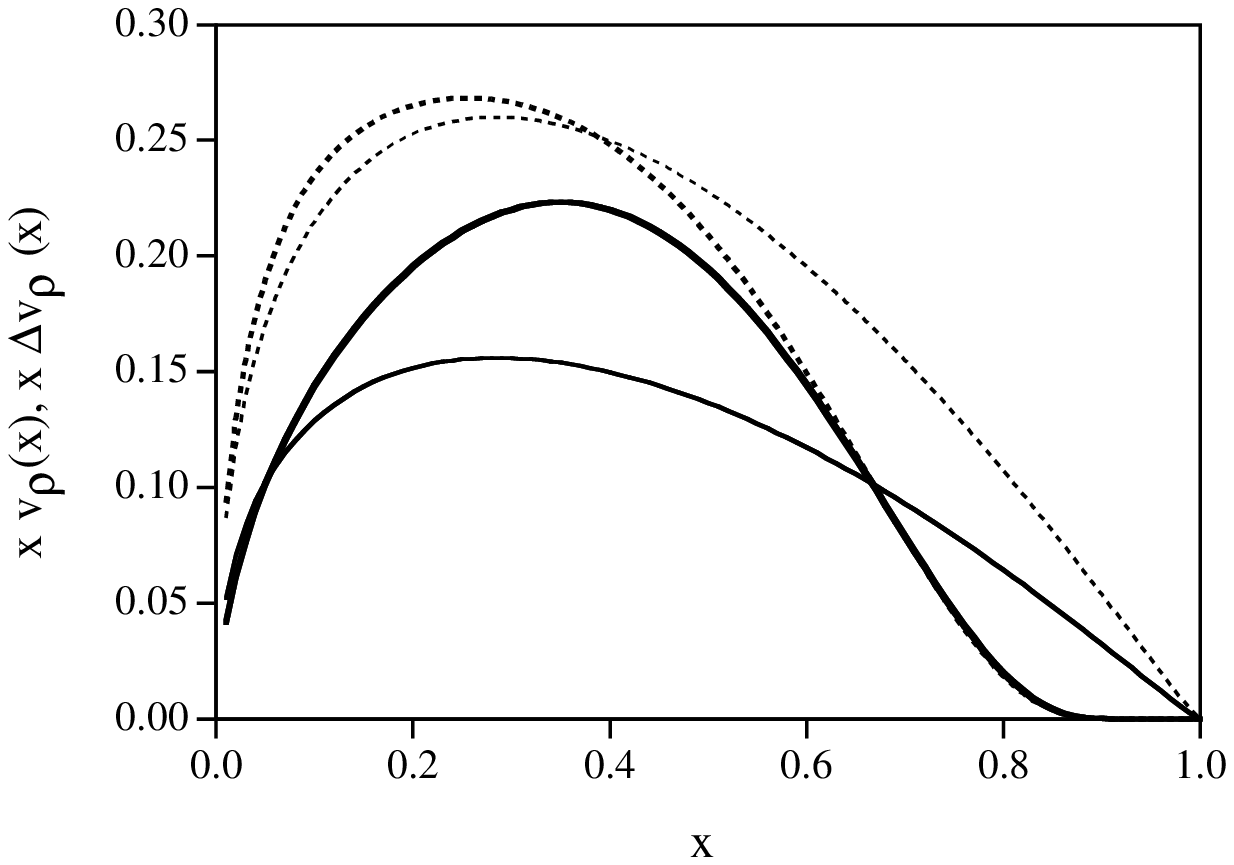,height=10cm,width=6cm,clip=,angle=0} \nonumber
&~~~&
\epsfig{file=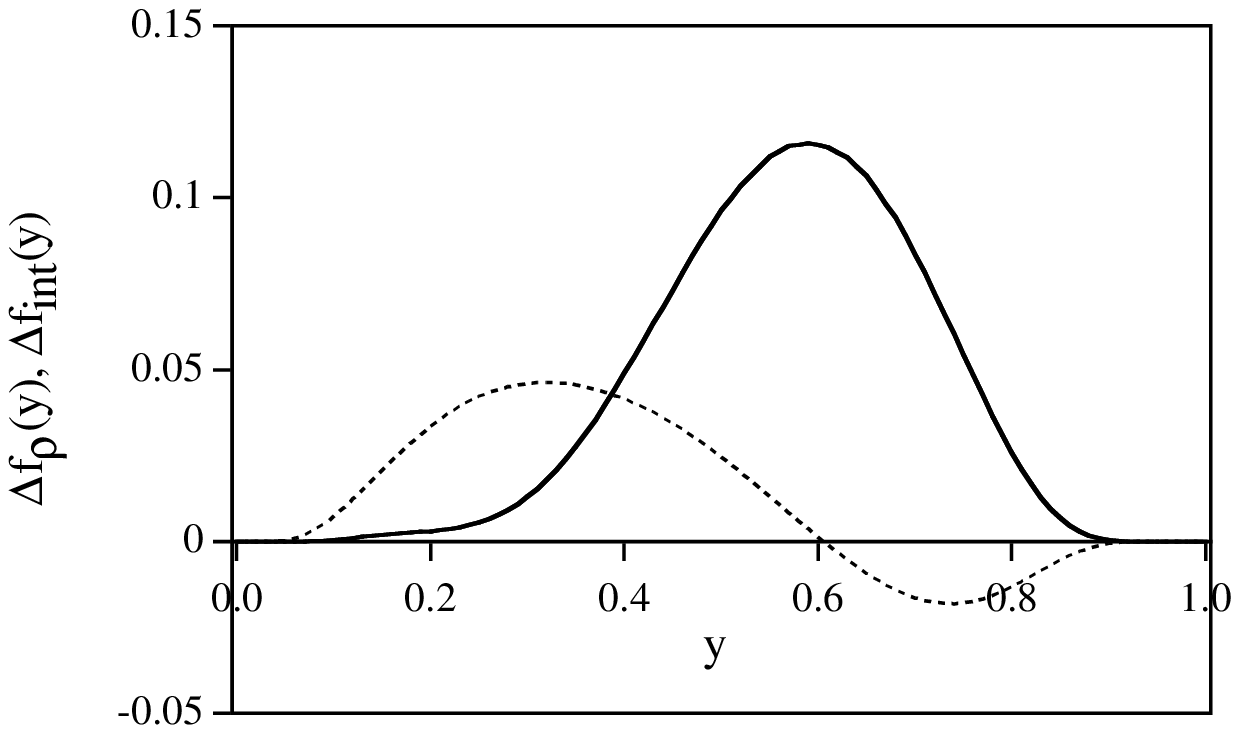,height=10cm,width=6cm,clip=,angle=0} \nonumber
\eea
\vspace{-7.5cm}
\bea
\parbox{5.7cm}
{\footnotesize
Figure 3: The polarized and unpolarized valence quark distributions of
the $\rho$ meson at $Q^2=4$~GeV$^2$.
See the text for more explainations.}
\hspace{0.5cm}
\parbox{5.7cm}
{\footnotesize
Figure 4: The polarized fluctuation functions $\Delta f_\rho$ (the solid curve)
and the interference term $\Delta f_{int}$ (the dashed curve).
$\Lambda_{oct}=1.08$~GeV, $\Lambda_{dec}=0.98$~GeV.} \nonumber
\eea
\end{center}                                                       
\end{figure}

We adopt two prescriptions for $\Delta v_\rho$ (see Fig. 3):
(i) employing the MIT bag model calculation,
$\Delta v_{\rho}^{MIT}(x)$ (the thick solid curve) 
and (ii) adopting the ansatz $\Delta v_\rho(x)=0.6 v_\pi(x)$ (the thin solid curve)
as in \cite{RFriesS}.
The parameters of the MIT bag model calculation are fixed by fitting the calculated
unpolarized parton distribution of the $\rho$ meson (the thick dashed curve)
to the Gluck-Reya-Schienbein parameterization \cite{GRS99}
for the valence parton distribution of the pion (the thin dashed curve).
The first moment of $\Delta v_{\rho}^{MIT}(x)$ is found to be about $0.60$ at 
$Q^2=4$~GeV$^2$, which is in agreement with the lattice value \cite{Lattice} of $0.60$.
It can be seen that the distribution $ 0.6\,xv_\pi(x)$ is smaller
than $x \Delta v_{\rho}^{MIT}(x)$ in the intermediate $x$ region,
although both distributions satisfy the same normalization condition.
Also the bag model calculated parton distribution has a different
$x$-dependence from the unpolarized distribution.

The fluctuation functions $\Delta f_\rho$ and $\Delta f_{int}$
in Eq.(\ref{xDeltadu})
are calculated by using time-ordered perturbation theory
in the infinite momentum frame \cite{FGCaoS,HHoltmannSS}.
(see Fig. 4. $\Lambda$ is a cut-off parameter in
the phenomenological form factor introduced to describe
the unknown dynamics in the fluctuation $N\ra MB$.)
It can be seen that the maximum of $\Delta f_{int}$ is about $40\%$ that of
$\Delta f_\rho$.
So the interfence contribution
to $x(\Delta \dbar -\Delta \ubar)$
will not be negligible, although
$\Delta f_{int}$ changes sign from positive to negative at about $y=0.6$.

The results for $x(\Delta\dbar -\Delta \ubar)$ are shown in Fig.~5.
\begin{figure}
  \begin{center}
\epsfig{file=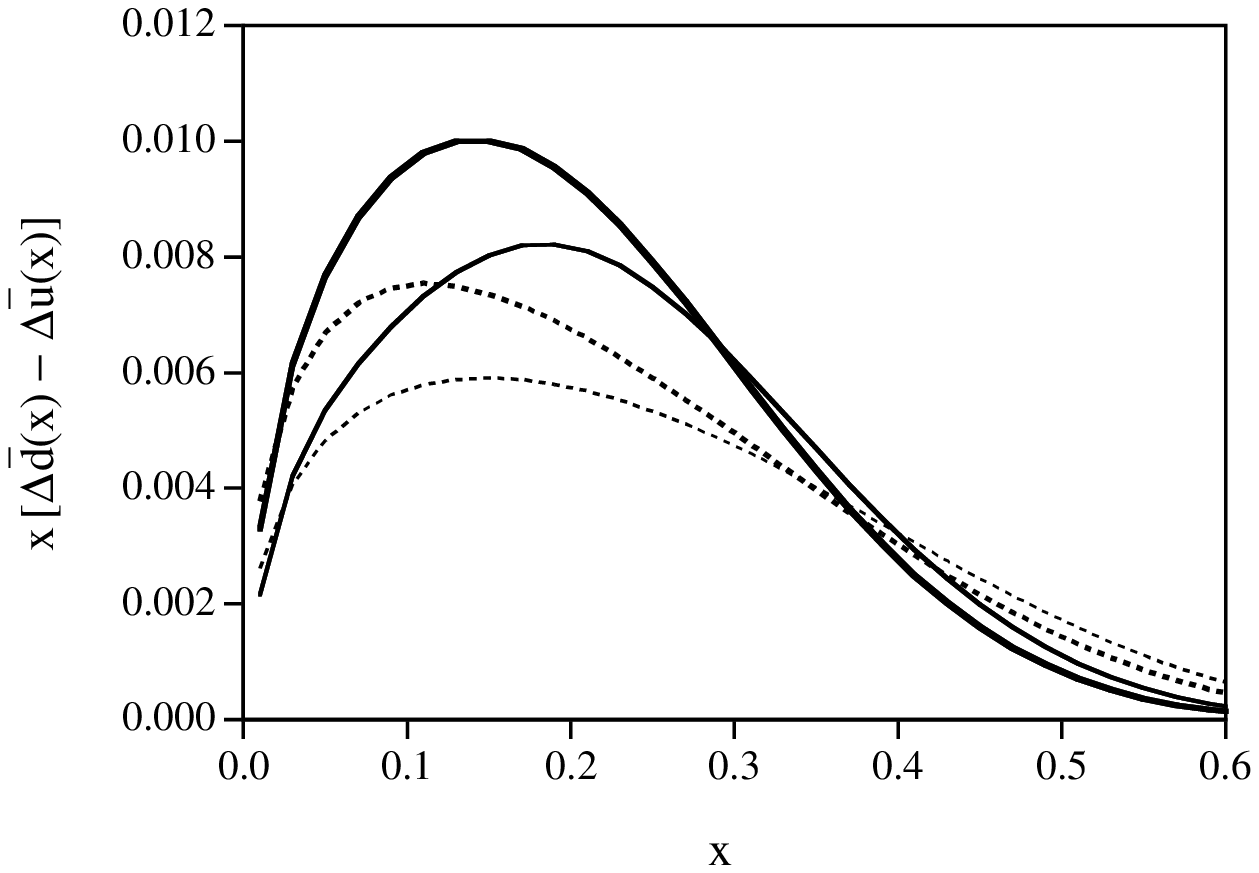,height=9cm,width=6cm,clip=,angle=0}
  \end{center}  
  \vspace{-5.7cm}                                                       
{\footnotesize Figure 5: The flavour asymmetry of the anti-quark in the proton.
The solid curves are the predictions using $x\Delta v^{MIT}_\rho$, while
the dashed curves are obtained by using $ 0.6 \, xv_\pi(x)$.
The thin curves are the results without interference contribution while the thick
curves are the results with interference contribution.}
\end{figure}
The interference effect increases sizably the predictions for the 
flavour asymmetry, and pushes the curves towards the small $x$ region 
due to  $\Delta f_{int}$ being peaked at smaller $y$ ($y_{max}\simeq 0.3$) 
than the $\Delta f_\rho$ ($y_{max}\simeq0.60$). 
Also the calculations with $x \Delta v_{\rho}^{MIT}(x)$ 
are larger than that with $ 0.6\, xv_\pi(x)$ 
in the intermediate $x$ region, and have their maxima at larger $x$.

The integral 
\bea
I_{\Delta}  &=&  \int_{0}^{1} dx [\Delta \dbar(x) -\Delta \ubar(x)] \nonumber \\
&=&  \int_{0}^{1} dx \Delta v_{\rho}(x) 
\int_{0}^{1} dy [\Delta f_{\rho}(y) + \Delta f_{int}(y)]
\eea 
will be the same for both models for the polarized parton distribution of the 
$\rho$ as they have the same first moment for the polarized distribution.
We find the integral to be $0.023$ ($0.031$) without (with) the interference 
terms for $\Lambda_{oct}=1.08$~GeV and $\Lambda_{dec}=0.98$~GeV.
The interference effect increases the integral by about $30\%$. 
The prediction for the integral $I_{\Delta}$ has a strong dependence on the cut-off
parameters $\Lambda_{oct}$ and $\Lambda_{dec}$. For example,
the results with (without) interference contribution vary from $0.0043$ ($0.0027$)
to $0.033$ ($0.020$) for the cut-off parameters changing from
$\Lambda_{oct}=\Lambda_{dec}=0.8$~GeV to $1.10$~GeV.
Clearly these values obtained using the meson cloud model are very different 
from those obtained using the chiral quark-soliton model \cite{LargeNC}
which have a magnitude of around $0.3$.
It is interesting that both models agree well with the experimental data 
for the unpolarized asymmetry, yet predict very different results for the 
polarized asymmetry. As the magnitude of the predicted polarized asymmetry 
appears to be fairly natural in each of these models, experimental data will 
provide a valuable test of these models, and give insight into the relation 
between helicity dependent and helicity independent observables in 
quark models.

\subsection{FSB from `Pauli blocking'}

Now we considere the contribution to the asymmetry arising 
from `Pauli blocking' effects \cite{WMelnitchoukST,ASignalT,FieldF}. 
In a model such as the bag model, where the valence quarks are confined 
by a scalar field, the vacuum inside a hadron is different from the vacuum
outside. 
This manifests itself as a distortion in the negative energy Dirac sea, which
is full for the outside (or free) vacuum, whereas there will be empty states in 
the Dirac sea of the bag.
To an external probe this change in vacuum structure appears as an 
intrinsic, non-perturbative sea of $q\bar{q}$ pairs. 
This change in the vacuum is similar to the change in the Fermi-Dirac distribution 
when the temperature is raised above absolute zero. 
Now when a quark is put into the ground state of the bag it wil have the effect of 
filling some of the empty negative energy states in the sea of the bag vacuum. 
The reason for this is that the ground state wavefunction can be written as a 
wavepacket in terms of plane wave states of positive and negative energy, with the 
energy distribution of the wavepacket centred at the ground state energy eigenvalue, 
but with non-zero contributions from negative energy plane waves. 
Hence the presence of a quark in the bag ground state lowers the probability of 
a negative energy state being empty, which is the same as lowering the probability 
of finding a positive energy antiquark. 
As the proton consists of two up quarks and one down quark, the probability of 
finding a $\bar{u}$ antiquark is reduced more than the probability of finding a 
$\bar{d}$ antiquark {\it i.e.} $\bar{d} > \bar{u}$. 

When we include spin in the analysis of Pauli blocking, we find that putting a 
spin up quark into the bag ground state has the effect of filling some of the 
negative energy spin up quark states in the bag vacuum, which is equivalent to 
lowering the probability of finding a positive energy spin down antiquark. 
As the $SU(6)$ wavefunction of the spin up proton is dominated by terms with the 
two up quarks having spin parallel to the proton spin and the down quark having 
spin anti-parallel, Pauli blocking predicts that the probabilities of finding spin 
down $\bar{u}$ antiquarks and spin up $\bar{d}$ antiquarks are reduced {\it i.e.} 
$\bar{u}^{\ua} > \bar{u}^{\da}, \; \bar{d}^{\da} > \bar{d}^{\ua}$ or 
$\Delta \bar{u}(x) \geq 0, \; \Delta \bar{d}(x) \leq 0.$

We estimate the contribution of the Pauli blocking effect to the 
polarized asymmetry using the Adelaide group's argument for calculating 
parton distributions in the bag model. It was found
\bea
\bar{d}(x) - \bar{u}(x) =  F_{(4)}(x), ~~~
\Delta \bar{d}(x) - \Delta \bar{u}(x) =-\frac{5}{3} G_{(4)}(x).
\eea
where $F_{(4)}(x)$ and $G_{(4)}(x)$ are the spin independent and spin dependent 
kinematic integrals over the momentum of the intermediate four quark state.
As $G_{(4)}(x)$ is positive at all $x$, Pauli blocking gives a negative 
contribution to the spin dependent flavour asymmetry in the sea, whereas the 
meson cloud contribution tended to be positive.
Also noting that as $F_{(4)}(x) \geq G_{(4)}(x)$, we can integrate over all $x$ 
and then obtain an upper limit for the size of the Pauli blocking contribution to the spin 
dependent asymmetry in terms of the contribution to the spin independent asymmetry:
\bea
-\int_{0}^{1} dx [ \Delta \bar{d}(x) - \Delta \bar{u}(x) ] \leq 
\frac{5}{3} \int_{0}^{1} dx [ \bar{d}(x) - \bar{u}(x) ]. \label{pbsums}
\eea
As an estimate for the integral on the rhs of Eq.~(\ref{pbsums}) we may 
use the value of 0.07 given by the analsis of reference \cite{WMelnitchoukST}.
This then gives an upper limit of about 0.12 for the magnitude of the integral 
over the polarized asymmetry. 
In the bag model, the ratio $G_{(4)}(x) / F_{(4)}(x)$ varies from about $0.8$ at 
low $x$ to unity at large $x$, which gives us a value of about $-0.09$ for the 
integrated polarized asymmetry.
While these values are calculated at some scale appropriate to the bag model, 
the values of the integrals are not much affected by evolution up to experimental 
scales, so we expect the relation between polarized and unpolarized sea 
asymmetries to be approximately correct at all scales.
The value of the Pauli blocking contribution to the integrated polarized asymmetry 
is much larger than that we have calculated in the meson cloud model, 
in contrast to approximate equality in the unpolarized case. 
Thus the experimental observation of any asymmetry in the polarized sea 
distributions is much more a test of the Pauli blocking hypothesis than of the 
meson cloud model.
We estimate that the contribution to the Bjorken sum rule from Pauli blocking
plus meson cloud effects is about 5-10\% of the value of the sum rule.


\section{Summary}
We report a study on the flavour asymmetry of
the non-strange polarized 
anti-quarks using the meson cloud model and `Pauli blocking'.
In the meson cloud model,
we have included the contributions from both the vector meson cloud and
the interference terms between pseudoscalar and vector mesons.
It was found that the interference terms can provide sizable contribute
to the asymmetry in the intermediate $x$ region.
We have also discussed the effect of `Pauli blocking' on the asymmetry, and have 
seen that this effect gives a larger contribution to the asymmetry than meson 
cloud effects, in contrast to the unpolarized case.

\section*{Acknowledgments}
We would like to thank Thomas, Melnitchouk, Schreiber and
Tsushima for helpful discussions.
This work was partially supported by the Science and Technology Postdoctoral
Fellowship of the Foundation for Research Science and Technology, and the 
Marsden Fund of the Royal Society of New Zealand.

\end{document}